# A 3D Finite Element model of the human face for simulation in plastic and maxillo-facial surgery.

M. Chabanas and Y. Payan
TIMC Laboratory , University of Grenoble, France

ABSTRACT

This paper introduces a new Finite Element biomechanical model of the human face, which has been developed to be integrated into a simulator for plastic and maxillo-facial surgery. The idea is to be able to predict, from an aesthetic and functional point of view, the deformations of a patient face, resulting from repositioning of the maxillary and mandibular bone structures. This work will complete the simulator for bone-repositioning diagnosis that has been developed by the laboratory. After a description of our research project context, each step of the modeling is precisely described: the continuous and elastic structure of the skin tissues, the orthotropic muscular fibers and their insertions points, and the functional model of force generation. First results of face deformations due to muscles activations are presented. They are qualitatively compared to the functional studies provided by the literature on face muscles roles and actions.

## 1. INTRODUCTION

Orthognathic surgery addresses skeletal deformities of the craniofacial region in the sagittal, coronal and transverse facial planes.  The correction of these deformities requires the repositioning of facial bones and their dental structures.  In this framework, detailed pre-surgical planning has to be carried out to accurately determine the exact amount of movement for the jawbones and teeth. Moreover, considerations about the aesthetics of the face and the functionality of the facial muscles have to be taken into account, to try to predict the « external aspect » of the patient face, following the repositioning of the bone structures. This point is crucial from the patient point of view, but unfortunately lacks in the simulations that are driven nowadays.

This paper addresses this last point, with the introduction of a new 3D biomechanical model of the human face and its coupling with a simulator for the repositioning of cranio-facial bone structures.

## 2. RESEARCH CONTEXT

In the course of the European Image Guided Orthopaedic Surgery (IGOS) project, the CAMI group (Computer Assisted Medical Intervention) of TIMC Laboratory developed a prototype for a 3D simulator for bone-repositioning diagnosis during maxillo-facial surgery (Bettega et al., 2000 (a); Bettega et al., 2000 (b)). The main idea was to provide the surgeon with a tool that offers the possibility, from patient CT scans, to plan the maxillary and mandibular osteotomies, taking into account cephalometric and orthodontic constraints. For this, a 3D virtual model of the patient skull is reconstructed and interactively used as a basis for the definition of a new 3D cephalometry and for the simulation of osteotomies. First tests were carried out on a dry skull and issues in the efficiency of this simulator were discussed (Bettega et al., 2000 (a)).

The aim of this paper is to introduce a new biomechanical model of the human face that would be coupled to the software developed in the IGOS project framework, in order to offer a complete simulator for plastic and maxillo-facial surgery. To do this, our strategy will be divided into three steps:

(1) A "standard" 3D model of the human face will be developed, with focuses on its anatomical, functional and physical realism. This step is the main topic of this paper.
(2) Starting from patient data (CT scans and MRI), our "Mesh-Matching" algorithm (Couteau et al., 2000) will be used to adapt the "standard" model to the patient morphology, thus generating a Finite Element model of the patient face.
(3) This 3D model of the patient face will be attached to the virtual reconstruction of the patient skull and used to predict facial aesthetics and behavior, after bone repositioning. This step will not be described in this paper and concerns current research.

3. BIOMECHANICAL FACE MODEL

3.1 Face anatomy

Facial skin has a layered structure composed of the epidermis (a superficial 0.1 mm thick layer of dead cells), dermis (0.5 - 3.5 mm thick) and hypodermis (fatty tissues connected to the skull). Many facial muscles are inserted between those skin layers and the underlying bone structure. In the case of maxillo-facial surgery, the great majority of interventions act on the upper and lower maxilla. This is the reason why our main focus for skin face modeling was dedicated to the lower part of the face, with a special emphasis on soft tissues surrounding human lips.

More than ten muscles act on lips shape, the great majority of them being pair muscles along the sagittal plane. Most of them are dilators (with a distended action like skeletal muscles), and are gathered around the lips. They all have the same kind of insertions: one into skull bone, and the other one inside the lips, onto a constrictor muscle which fibers run around the lips: the orbicularis oris (figure 1).

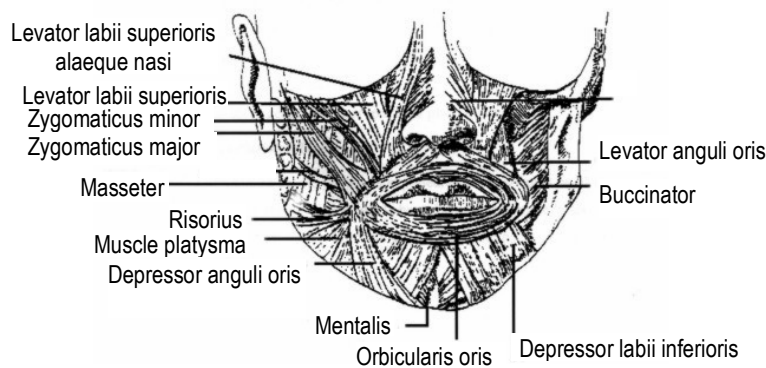

**Figure 1 :** Face anatomy (from Bouchet & Cuilleret, 1972)

3.2 Face modeling

The first developed models of the human face focused on computer animation and were motivated by a need for external realism, from a behavioral point of view. Their modeling was mainly based on discrete mass-spring structures, regularly assembled inside facial tissue. In a first step, those kind of discrete models were introduced in the framework of computer assisted maxillo-facial surgery (Waters, 1996, Keeve et al., 1996 (a)). Then, arguing that a precise modeling of soft tissues deformation needs a continuous description, Finite Element (FE) models were developed (Keeve et al., 1996 (b); Keeve et al., 1998; Koch et al., 1998). Those models were based on an automatic 3D meshing of the FE structure, on an isotropic description of tissues mechanics, and proposed, for some of them, principles for the modeling of muscular actions (Koch et al., 1998).

As we stated earlier, our strategy was to build a 3D face model that specifically addresses the simulation in the domain of plastic and maxillo-facial surgery, with a modeling framework that offers the possibility:

(1) to passively predict the face aesthetics after bone repositioning,

(2) to estimate the functional deformations of the face under muscles actions.

This last point is particularly important to ensure for example a symmetrical smiling or to guaranty no perturbation for speech production.

The first objective will induce a continuous modeling of facial tissues. For this, the partial differential equations of the linear elasticity theory were discretized through the Finite Element Method. A particular attention will be given to the definition of the 3D FE mesh, that will have to be sufficiently adapted to face anatomy: symmetry along the sagittal plane, precisions near crucial anatomical structures (nose and lips), and possibility to include muscular structures.

The second objective will need a precise modeling of muscular fibers, with a description of their course and insertion points, and with specific mechanical properties. From this point of view, and as a consequence of the complex interweaving of face muscles, we have chosen to limit the description of the deformations to the linear elasticity modeling. However, we offer the possibility to model the anatomical specificity of the muscles, i.e. their orthotropic structure (i.e. muscles reinforced by fibers).

*3.2.1. 3D Finite Element Mesh*

We started from a 3D mesh surface of face skin (Guiard-Marigny *et al.*, 1996). This mesh was slightly modified and used to build two other surfaces, in order to model a volumetric mesh (figures 2). The first one is just scaled from the skin mesh, while the other one corresponds to its projection onto a standard skull. Points of this third surface are rigidly fixed, except those located around the mouth and inside the cheeks which are not fixed to the skull. A volumetric mesh, composed of two layers of material, was thus built, modeling the collagen and fatty tissues of the skin structure.

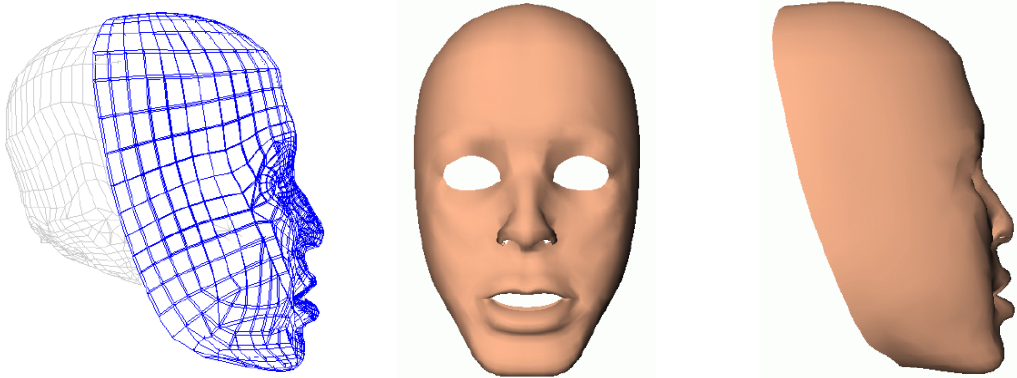

**Figure 2 :** Finite Element mesh (left) and rendered views (middle and right)

Muscles were defined by making some "holes" inside this global 3D mesh, and by inserting other FE meshes with elements representing the body of the muscles. The four main muscles acting on lips were inserted, with elements running in the superficial thickness of the mesh in the case of the orbicularis oris, or elements running from the extremity of the muscle, attached to the skull, to the other insertion points, fixed onto the orbicularis muscle (in the case of the zygomaticus, risorius and depressor labii muscles). Figures 4 (left panels) plot the final 3D mesh, with the areas (dashed elements) associated to FE models of the muscles.

### 3.2.2. Mechanical characteristics

As a starting point, the global face mesh, mainly describing passive collagen and fatty tissues, was modeled as an isotropic linear FE structure. The biomechanical characteristics of this structure were chosen in order to model skin tissues quasi-incompressibility and to replicate mechanical measurements made on skin. Tissues incompressibility was modeled with a Poisson's ratio value close to 0,5. Young's modulus value was chosen to match the measurements reported by Fung (1993). Those measurements were made on rabbit skin, but are, from our knowledge, the only reference about skin mechanical characteristics (they are, from this point of view, used by other authors in FE modeling (Keeve et al., 1998)). The value that was retained in our simulations corresponds to the small-deformation slope of the skin stress-strain relationship reported by Fung, i.e. 15 kPa.

Concerning muscles mechanical characteristics, we took the values reported by Duck (1990): 6.2 kPa for a human muscle in its rest position, and 110 kPa for the same muscle when it is contracted. Those values are coherent with measurements reported by Ohayon et al. (1999) on cardiac muscles, considered as dense muscles: 30 kPa at rest and 300 kPa when activated. Moreover, due to the orthotropy of muscular fibers, we raised the Young modulus value to 110 kPa only for the modulus corresponding to the principal direction of the muscle. The Young modulus corresponding to the two orthogonal directions were maintained at the 6.2 kPa value, even when muscle is activated. Finally, as for the passive elements associated to the rest of the face, Poisson's ratio value was chosen close to 0,5, considering that muscles are mainly composed of water and thus quasi-incompressible.

## 4. SIMULATIONS

First simulations were carried out on the standard model to validate the deformations of the face under muscles actions, according to functional studies (Hardcastle, 1976). Each muscle was activated through a force applied in the direction of the fibers, with a level close to the measurements reported by Bunton and Weismer (1994) on human tongue and lips (some Newtons in general). The Young modulus value associated to the main direction of the fiber was linearly raised with force activation, between 6.2 kPa (no activation) and 110 kPa (maximal activation). The orientations of each muscle fibers are not straight but on the contrary show a more or less important curvature among muscles. Therefore, a functional model of distributed force was adopted to simulate activation: the force is distributed on each node of the muscle FE structure, with a value proportional to the curvature. Figure 3 (left panel) shows this distribution for the zygomaticus muscle: the global force is applied to the extremity of the muscle, but residual forces, due to the distribution as a function of the curvature, are applied to nodes belonging to the body of the muscle, with a resulting action that will try to re-align body muscles elements. The right panel of figure 3 plots force distribution for the orbicularis muscle. This model is particular as it was considered as a sphincter muscle, without any extremity. Anyway, our functional distributed model allows the generation of forces on each node of the muscle, with an amplitude function of the curvature: therefore the idea is that an equilibrium can be reached if the sphincter muscle shape is close to a circle, with a curvature constant among nodes.

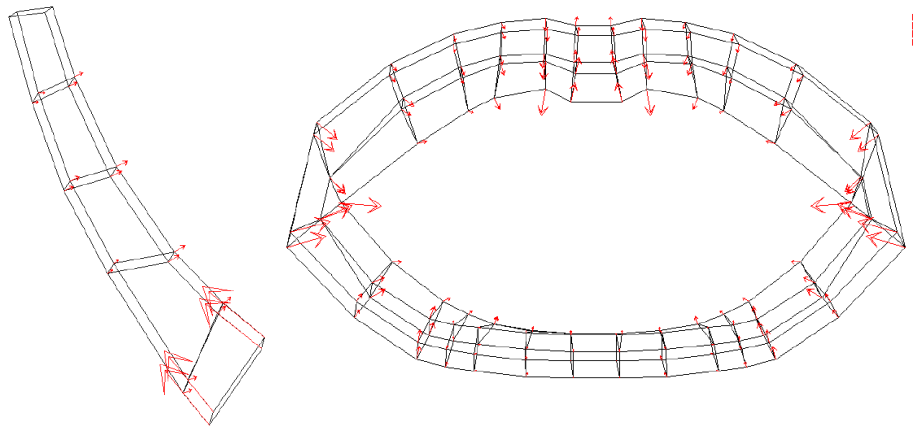

**Figure 3 :** Forces repartition on mesh nodes
Zygomaticus major (left) and Orbicularis oris (right)

Figures 4 plot the face deformations induced by muscles. For each muscle activation, computation time is about 5 minutes on a 450Mhz PC (resolution of the FE partial differential elasticity equations provided by the *Castem 2000* package). As the model was defined in the scope of linear elasticity, those figures can be reduced thanks to pre-calculus (Cotin et al., 1996).

**Zygomaticus major**

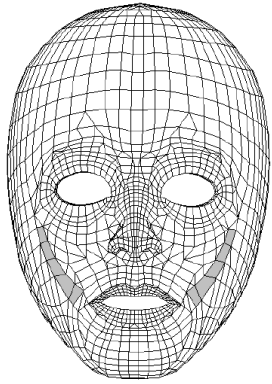 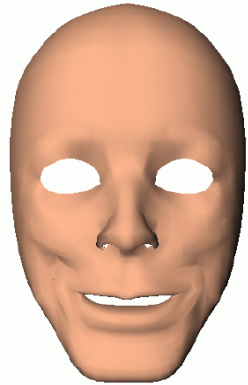 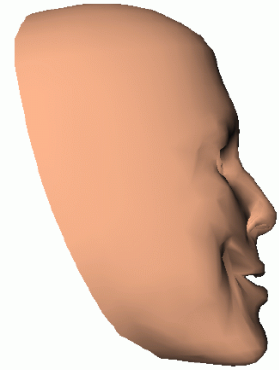

**Risorius muscle**

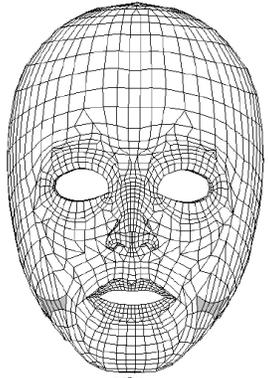 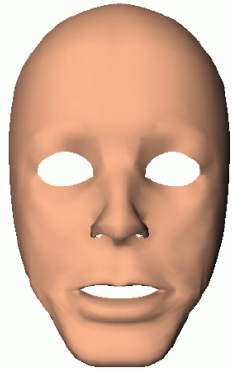 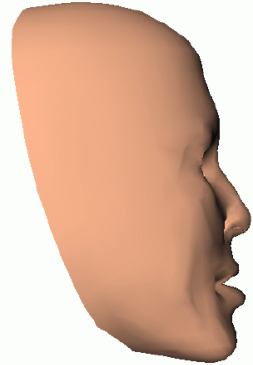

**Depressor labii inferioris muscle**

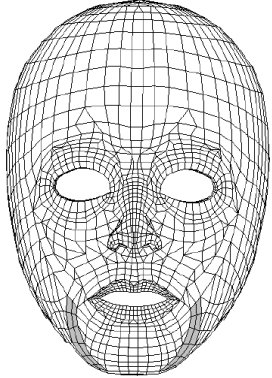 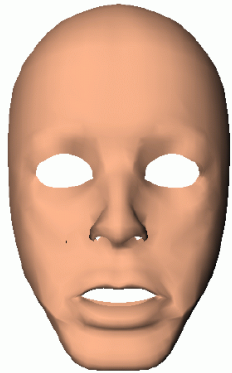 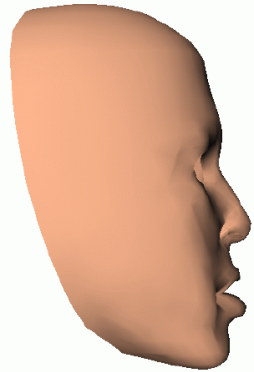

**Orbicularis oris muscle**

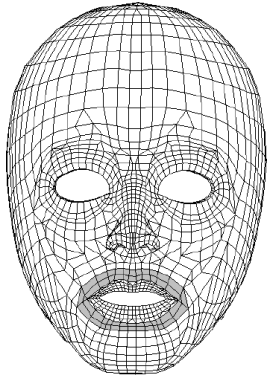 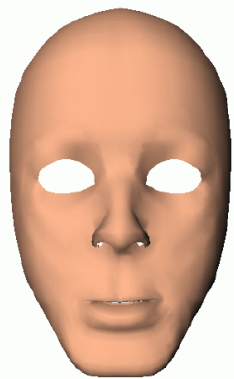 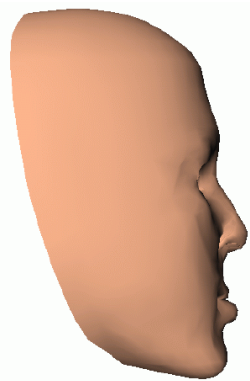

**Figure 4 :** Muscles modeling : geometry and actions
Left panel : muscle FE mesh (dashed elements)
Middle panel : induced deformations (face)
Right panel : induced deformations (profile)

## 5. DISCUSSION AND PERSPECTIVE

Those simulations qualitatively validate the modeling, as the face deformations induced by each muscle activation are coherent with the functional studies provided on face muscles roles and actions (Hardcastle, 1976). The zygomaticus major muscle pulls the lip corners upwards, the risorius muscle stretches the lips, while the depressor labii muscle lows lip corners. Concerning the activation of the orbicularis muscle, a lip closure gesture can be observed, while a slight backward movement of the upper lip can be seen on the profile view. This movement should not be observed once a geometric model of the teeth (with no penetration constraints for the lips) will be added to the face model.

As a perspective, the next step of our research project will consist in adding layers to complete the model, in order to include other face muscles (both superficial and deeper ones: levators and depressors, masseter, etc.) and skin epidermis. Then, quantitative validation of the modeling will have to be carried out, on a "normal" subject. The idea will be first, to measure mechanical properties of the patient (with indentation experiments for example) as well as physiological muscle properties (electromyography), and second, to confront deformations simulated by the model, with skin deformations observed on the subject. Finally, the global face model will be integrated into the simulator for bone structures repositioning, and quantitatively validated on patients (confrontations between aesthetics and functionality predicted by the model, and observations made on the patient, after surgery).